# On a Distributed Approach for Density-based Clustering

Nhien An Le Khac
School of Computer Science
University College Dublin
Belfield, Dublin 4, Ireland
email: an.lekhac@ucd.ie

M-Tahar Kechadi
School of Computer Science
University College Dublin
Belfield, Dublin 4, Ireland
e-mail: tahar.kechadi@ucd.ie

*Abstract* — **Efficient extraction of useful knowledge from these data is still a challenge, mainly when the data is distributed, heterogeneous and of different quality depending on its corresponding local infrastructure. To reduce the overhead cost, most of the existing distributed clustering approaches generate global models by aggregating local results obtained on each individual node. The complexity and quality of solutions depend highly on the quality of the aggregation. In this respect, we proposed for distributed density-based clustering that both reduces the communication overheads due to the data exchange and improves the quality of the global models by considering the shapes of local clusters. From preliminary results we show that this algorithm is very promising.**

## I. INTRODUCTION

Nowadays, data mining (DM) [1] is used nearly in every field where gathered data is available in abundance. For in instance, in market analysis, the data can be used to analyse client behaviour or optimise production and sales, in climatology, to extract information of satellite observations, in biology, to analyse large amount of genomic data and in many other fields. In order to cope with gigabytes or even terabytes of data, a natural step is to use the power of parallel and distributed machines, and there as been parallel versions for centre-based DM algorithms [2]. These parallel and distributed machines were mostly clusters of computers or Grids [3]. In this case, large amounts of datasets are divided (either horizontal or vertical) into disjoint partitions and then scattered on computing nodes. DM algorithms are carried out on each node to create local models. These local models will be aggregated to build global models. However, in the case of a very large dataset, the local models are still large to be sent over the network, which will create significant communication overheads.

Meanwhile, today, massive amounts of data are stored in different sites (nodes) as they were produced. In this context, distributed data mining (DDM) techniques have become a necessity for analysing these large and multi-dimensional datasets. DDM is more appropriate for large-scale distributed platforms where datasets are often geographically distributed and owned by different organisations. Many *DDM* tasks such as distributed association rules and distributed classification [4–6][17–19] have been proposed and developed in the last few years. However, only a few research concerns distributed clustering for analysing large, heterogeneous and distributed datasets. Sending a huge datasets over the network for mining and management is a performance issue due to the limitation of bandwidth, network bottleneck, etc. Therefore, recent researches [7][8][14] have proposed a distributed clustering model that is based on two key steps: perform partial analysis on local data at individual nodes (sites) and then send the obtained models to a central node to generate global models by aggregating the local results. One of the biggest challenges is to build good quality global models, as local models do not contain enough information for the merging process. In other words, there is always a trade-off between the size and the quality of the representatives in local models. Usually, a small number of representatives are required in order to reduce the communication cost. However, these representatives could not reflect all-important features of their clusters.

Besides, global models of current approaches depend on local clustering techniques. For instance, if the local model is created by density-based clustering techniques then the global model is also built on the same paradigm. Moreover, the shape of a cluster created is also important because it can show exactly the trend of data objects especially for spatial data. Therefore, good representatives of local models should preserve the shape of clusters as one of the important information.

In this paper, we propose a new approach of distributed clustering. In this approach, local clustering is also carried out at each node to build local models. These models are aggregated to generate global ones. So far the approach is similar to ones mentioned above, but our approach introduces new concepts of characterising the local models. For instance, local models are characterised not only with traditional representatives but we also added their boundaries. The merging of local model is also based on these boundaries. Therefore, local clustering can be any clustering types such as centre-based, density-based, etc. Moreover, in our approach, we present different regenerating methods that rebuild local datasets on the server side. This feature can help to increase the quality of global models. In the following we will focus on the algorithms for creating the local models as well as regenerating the clusters.

The rest of this paper is organized as follows: Section II deals with background and related projects then we will present and discuss our new distributed clustering approach in Section III. Section IV presents our preliminary evaluations of core algorithms and analysis. Finally, we conclude in Section V.

## II. BACKGROUND

In this section, we briefly review the distributed data mining. We then study different approaches of distributed clustering.

### A. Distributed data mining

In a *DM* application, for which the datasets are stored in a central data repository, the process is normally called centralised DM (CDM) or local DM. This approach is appropriate for small datasets. CDM is not suitable for exploring very large and distributed datasets because of the limited computing resources, privacy and security, etc. In a distributed environment, the data may be distributed among different locations for various reasons: an application by its nature is distributed or sometimes the data are distributed for better scalability and disk space management. The centralised DM techniques do not consider all the issues of data-driven applications such as scalability, distribution and heterogeneity. DDM approaches have been proposed as a good alternative. The type of DDM techniques, we are interested on, would able to learn new models from distributed data without exchanging the raw data.

A DDM system normally includes components such as: data pre-processing, mining algorithms, communication subsystems, resource management, user interfaces, etc. The main goal of a DDM system is to provide an environment for accessing distributed data, allocating resources, monitoring the entire mining process and interpreting the results. A DDM system should offer a flexible environment to adapt various kinds of applications. In the early stages, small distributed systems such as clusters of workstations have been exploited to deploy DDM applications. Today, current *DM* problems require new infrastructures and architectures to face new challenges; large volumes, heterogeneity, and complexity of the datasets.

### B. Distributed clustering

Clustering is one of the basic tasks in the data mining area. Basically, clustering enables to group data objects based on information that describes the objects and their relationships. The goal is to optimise similarity within a cluster and the dissimilarities between clusters. Most distributed clustering algorithms are based on a parallel approach and are normally well suited for homogeneous distributed data [9]. These algorithms are further classified into two sub-categories. The first consists of methods requiring multiple rounds of message passing. These methods require significant synchronization mechanisms. The second sub-category consists of methods that build local clustering models followed by the aggregation phase. These methods require only a single pass of message passing, hence, modest synchronisation requirements.

In [7], authors used DBSCAN [13] as a local clustering algorithm. They extended primitive elements of this algorithms such as core points, $\varepsilon$, *Minpts* by adding new concepts as specific core points, specific $\varepsilon$ to build a local representative at each site. The global model will be rebuilt by executing the DBSCAN algorithm on a set of local representatives with two global values: $Minpts_{global}$ and $\varepsilon_{global}$. $Minpts_{global}$ is a function of two local parameters i.e. $Minpts_{global}$ = 2 x *Minpts*. $\varepsilon_{global}$ is tuneable by the user and its default value is the maximum value of all $\varepsilon$ values of all local representatives.

[8] is also based on DBSCAN as a local clustering algorithm. This is an improvement of the previous approach where absolute core points were applied instead of specific core points. This approach also takes into account noise objects in its local models. Finally a hierarchical agglomeration is applied to aggregate all local models to build global ones. Another approach presented in [14] also applied a merging of local models to create global models. As mentioned in Section I, current approaches only focus on either merging local models or mining on set of local models to build global ones. If the local models cannot effectively represent local datasets then global model accuracy will be very poor.

## III. DISTRIBUTED CLUSTERING

In this section, we describe firstly a new distributed clustering model where the local models are based on the boundaries of clusters. We also present and discuss two different strategies for merging clusters.

Our model includes three main steps. In the first step, we cluster the datasets located on each local node and select good local representatives. Next, we send all the local models to the server. We regenerate data objects on the server basing on the local model representatives. The purpose of this step is to improve the quality of the global model regarding the small size of local models that do not have enough important information.

### A. Local model

Local model includes representatives created at each local site in the system. Normally, the clustering task creates the local representatives in most current approach [7][8][14]. Furthermore, in our approach, we focus on the shape of the clusters created because it is also an important feature as mentioned in Section I. Hence, we also take into account the boundaries of clusters as a part of local representatives. To calculate the boundaries of a cluster we develop an algorithm called a balance vector algorithm. It attempts to detect clusters' boundaries and then use them as the main part of the local representatives. At a local site *i*, let $C_i$ be a set of clusters created, and $B_{Cj}$, $R_{Cj}$ be the boundary, the internal representatives of a cluster $c_j \in C_i$ respectively. The local model $L_i$ is defined by:

$$L_i = \{\bigcup_{j=1}^{n}\{B_{C_j} \cup R_{C_j}\} \cup P_i \mid B_{C_j} \in c_j, R_{C_j} \in c_j, c_j \in C_i\} \quad (1)$$

The set of boundaries of all clusters created at the site *i* are denoted by $B_i$ where $B_i = \cup B_{Cj}$.

$P_i$ is the local parameters applied to create local representatives. We also use these parameters in the merging process to build the global model. Hence, we add them into the local model $L_i$.

The internal representatives $R_{Cj}$ depend on the clustering algorithm applied. For instance, for the centre-based algorithm such as K-Means [99] or K-Medoids [88], $E_i$ can be a set of means or medoids objects. Meanwhile, if we focus on the shape and the density of clusters, then we can use the density-based algorithms such as DBSCAN [13], snnDBS [15]. In this case, $E_i$ can be a set of core objects [13] or even specific core points [7]. However, the selection of representative that can be added to the reduction set is still a challenge in terms of quality and size of this set. We can choose, for instance, medoids points, core points, or even specific core points [7] as representatives. Furthermore, the density of a cluster can also be represented by the number of data objects of a cluster or even by a mean density value and/ or by a set of density values describing the density in various areas of the cluster. The set of all internal representatives at the local site i can be noted by $R_i$ where $R_i = \cup R_{Cj}$.

The boundary $B_{Cj}$ of a cluster $c_j$ is the set of all boundary points in $c_j$. The boundary points are defined as points that outline the shape of a cluster. They can be seen as borders of clusters. Getting the borders is a difficult problem. This task has to work with clusters of different shapes that may contain holes, with different densities and it has to work with dataset in $n$ dimensions. The boundary points are visually characterised as those points that confine with a dense area on one hand and with an empty area on the other hand. In order to detect boundary points of a cluster, we develop the balance vector algorithm. This algorithm works with density-based clustering. Before explaining this algorithm in details, we define some basic concepts.

*1) ε-neighbourhood:* Given a cluster $C \subseteq \mathfrak{R}^n \equiv \{p_1, p_2..p_n\}$. The ε-neighbourhood $N_\varepsilon^C(p)$ of a point p in the cluster C is defined as the set of points $\{p'_i\} \in C$ so that the distance between to $p'_i$ and $p$ is less than or equal to $\varepsilon$.

$$N_\varepsilon^C(p) \equiv \{p'_i \in C \mid dist(p,p'_i) \leq \varepsilon\} \quad (2)$$

In order to check if a point $p$ is a boundary point, we determine in which direction the least dense area of $p$'s neighbourhood is located.

*2) Displacement vectors:* A displacement vector from $p'_i \in N_\varepsilon^C(p)$ to the point p is defined as:

$$\vec{v}_p = \sum_{p_i \in N_\varepsilon^C(p)} (p - p_i) \quad (3)$$

This vector points towards the area of the lowest density of the neighbourhood of $p$.

*3) Balance vectors:* Since we are concerned with the direction of the least dense area of the neighbourhood relative to the point p, the length of the vector does not hold any relevant information. We hence define the balance vector relative to the point p as:

$$\vec{b}_p = \begin{cases} \frac{1}{\|\vec{v}_p\|} \vec{v}_p & if \|\vec{v}_p\| > 0 \\ \vec{0} & otherwise \end{cases} \quad (4)$$

Fig.1 shows an example of a balance vector. The neighbourhoods of p are inside the circle, the balance vector is represented by the blue colour.

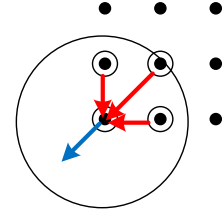

Figure 1. Example of a balance vector.

*4) Boundary points:* If a point is a boundary point, there should be no points towards the direction of the balance vector. This property allows us to separate boundary points and internal points. Normally, there are many options available to decide which empty area to look for in the direction of a balance vector. In our earliest approach, we check for an empty hyper-sphere whose centre lies on the line having the same direction as the balance vector (see Fig.3). This check can be formalised as follows:

$$Boundary(p) = \begin{cases} true & if (N_\varepsilon^C(p + \rho\vec{b}_b) = \phi \vee (\vec{b}_p \neq \vec{0}) \\ false & otherwise \end{cases} \quad (5)$$

However, this has proved to be far from ideal, as many points, easily identifiable as part of the boundary by the human, were in fact not identified, because other boundary points lied in the hyper-sphere. This is especially true when the point lies in a concave area of the boundary. We devised then a more accurate predicate, that, for each point p, checks for an empty area whose shape is the intersection of an hyper-cone of infinite height, vertex, axis $\vec{b}_p$ and aperture υ, where υ is a given angle.

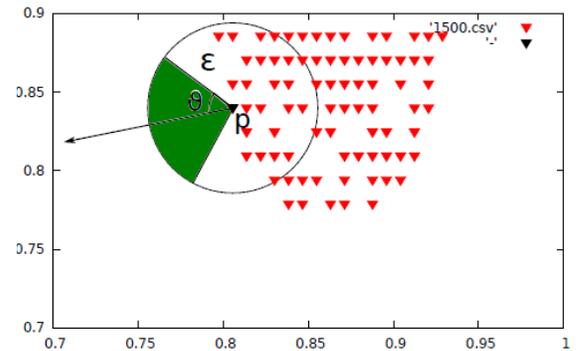

Figure 2. Boundary point check.

As shown in Fig.2, the area checked is highlighted in green. Formally, a boundary point is described as a Boolean predicate

$$Boundary(p) = \begin{cases} true & if\ \forall q \in N_\varepsilon^C(p), (q-p)\vec{b}_p < \cos(\upsilon) \\ false & otherwise \end{cases} \quad (6)$$

This condition is more efficient than the previous one (5). We discuss the comparison of these two conditions (5) and (6) as well as the parameter choice in Section IV. We can thus define the boundary $B_C$ of a cluster $C$ as the set of all boundary points in $C$ as follows:

$$B_C = \{p \in C: Boundary\ (p)\ is\ true\} \quad (7)$$

The algorithm for selecting the boundary points is described as in Table I.

Briefly, the boundary of clusters as well as their density information will construct the local model $L_i$ at the local site $i$ in the system.

In the next step, all local models from local sites are sent to the server in which global models will be built. Our model supports both synchronous and asynchronous communications.

TABLE I. BOUNDARY DETECTING ALGORITHM

| Input | Cluster $C$, set of balance vectors $\{\vec{b}_{p_i}\}$ for all points $p_i \in C$, parameter $\upsilon$. |
|---|---|
| Output | Boundary points $B_C$ of the cluster $C$ |
| 1: | $B_C \leftarrow C$ |
| 2: | **for all** *points p* in $B_C$ **do** |
| 3: |     **for all** *points q* in $N_\varepsilon^C(p)$ **do** |
| 4: |         **if** $(q-p)\vec{b}_p \geq \cos(\upsilon)$ **then** |
| 5: |             Discard $p$ from $B_C$ |
| 6: |             **break** |
| 7: |         **end if** |
| 8: |     **end for** |
| 9: | **end for** |

*B. Global model*

The last step is for merging local models. The merging process consists of two steps: boundary merging and regeneration. In the first step, boundaries of clusters from the local models are merged together by a boundary-based method. At the merging site, let $L_i$ be a local model received from the site $i$ and $B_i$ be the set of all boundaries in $L_i$. The global model $G$ is defined by:

$$G = \Gamma(\cup B_i),\ B_i \in L_i \quad (7)$$

where $\Gamma$ is a merging function. In our approach, we apply a modified version of the boundary detection algorithm (cf. Table I) on all $B_i$ collected from local sites as a merging function. Let $B$ be all $B_i$ collected, $B \equiv \cup B_i$. The inputs are $b_i \in B$, a set of balance vectors and $g\upsilon$. $g\upsilon$ is a global parameter and it is based on the value of all local parameters $\upsilon_i$ from $L_i$. The output of this algorithm is a global boundary set $GB$ that contains boundary of input objects i.e. all data objects in $B$. So, the global model $G$ contains the global boundary set $GB$.

Next, we carry out a regenerating process to add relevant objects into the global model $G$. Basing on the $GB$ set, we can regenerate these objects by using $R_i$ (cf. III.A). Let $C$, $B_C$ be a cluster and boundaries of $C$ respectively. Given a point $q$, $q$ is inside $B_C$ (*Inside(q,$B_C$)*) if:

$$(p_j - q) \bullet \vec{b}_{p_j} > 0,\ p_j\ \text{is the nearest neighbour of } q \text{ in } B_C \quad (8)$$

Meanwhile, the *minimal enclosing hypercube* of $B_C$ is defined by:

$$MEH(B_C)=\{x \in \mathfrak{R}^n |\ \forall p_i \in B_C: min(p_i) \leq x_i \leq max(p_i)\} \quad (9)$$

Given the cluster $C$'s average density $d_C$, we aim to build a cluster $C'$ that is similar to $C$ by generating points inside the minimal hypercube that encloses the points in the cluster boundary $B_C$, and only points that satisfy the condition (6) can be added to $C'$. This process should be iterated until the cluster $C'$ reaches the average density $d_C$. As there are many strategies to regenerate the points, we present the simplest one: random throw (Table II). In this strategy, the cardinality of $m = \|C\|$ of the cluster $C$ is used instead of $d_C$.

TABLE II. RANDOM THROW ALGORITHM

| Input | Cluster boundaries $B_C$, $m$. |
|---|---|
| Output | Cluster $C'$ |
| 1: | $C \leftarrow \varnothing$ |
| 2: | **while** $m > 0$ **do** |
| 3: |     $x \leftarrow Random(MEH(B_C))$ |
| 4: |     **if** *Inside(x, $B_C$)* **then** |
| 5: |         $C' \leftarrow C' \cup \{x\}$ |
| 6: |         $m \leftarrow m - 1$ |
| 7: |     **end if** |
| 8: | **end while** |

## IV. EVALUATION AND ANALYSIS

In this section, we present preliminary results of core algorithms described in the previous section. We evaluate our new model, its algorithms as well as their potential and improvement needed.

*A. Boundary Detection*

First, we test our *Boundary Detecting* algorithm on a 2-D dataset DS1 (Fig.3a). We also compare two detecting techniques: hyper-sphere (cf. (5) III.A.4) vs. hyper-cone (cf. (6) III.A.4). An important issue is the value of parameter $\rho$ (cf. (5), III.A.4). If it is too small, we cannot detect correctly the boundaries and if it is too large, it may delete boundary points because there are points in front of it. As shown in Fig.3b, the boundary points detected by sphere-based algorithm of clusters located at the upper-left corner do not clearly reflect the border of these clusters. The reason is these clusters gathered are near to each other. With a small value of $\rho$, many boundary points cannot be detected. When the value of $\rho$ is large enough, we miss some boundary points one near side of other cluster. In our case, the value of $\rho$ is selected to be the double of the mean distance to the furthest point in the neighbourhood. Meanwhile, the boundary points detected by cone-based algorithm can efficiently represent borders of these clusters (Fig.3c). However, when the clusters are isolated enough and with fewer concaves, sphere-based algorithm reports good results compared to cone-base one in terms of boundary points detected as shown in the rest two clusters in Fig.3. The hyper-cone algorithm requires two input parameters:

- The neighbourhood radius ε used to calculate the balance vector that co-indices with the radius of the hyper-sphere used to perform the boundary point check.
- The hyper-cone aperture υ.

By experience through experiments performed on 2-D and 3-D datasets – the best value of υ is around π/6. Besides, *ε* has a big impact on the results. It influences both two algorithms hyper-sphere and hyper-cone in three ways:

*1) Balance vector calculation:* A too small value can lead to vectors calculated on the basis of a very small numbers of neighbours, leading to vectors pointing in directions that are easily identifiable as "wrong". A similar problem occurs when too large values are selected, as the neighbourhood would not represent what can be understood as the local distribution of the data.

*2) Hole detection:* The algorithm is capable of representing holes in (empty areas inside) the cluster shape, and ε is proportional to the minimum size of the hole that we want to be rejected in the cluster outline. Too small values of ε can lead the algorithm to outline normal irregularities of the data as holes, identifying internal data points as boundaries. Pathologically small values allow the algorithm to identify almost all the data points in the cluster as boundary points. Too large values will instead ignore some holes that we would otherwise want rejected in the outline.

*3) Level of detail/Outline quality:* Small values tend to produce a good level of detail, when they are not pathologically small, and the algorithm identifies a lot of internal points as part of the boundary. Increasing *ε* tend to reduce the level of detail, eventually losing some features like the hook shape, and when a certain threshold is crossed when *ε* is close to the cluster radius - the whole cluster is considered as an hyper spherical blob, and only the most external cluster outline is extracted.

*B. Distributed clustering*

Next, we evaluate our distributed clustering model. The testing platform is a cluster of one server and three computing nodes. We also use Open-MPI [20] as the communication interface. In the first phase, our dataset is randomly divided in three equal partitions and then each partition is located on each node. We carry out then a local process on each node in the phase 2. This local process includes two steps: clustering by DBSCAN [13] to create set of local clusters and then detecting boundary of each cluster to build the local model (cf. III.A).

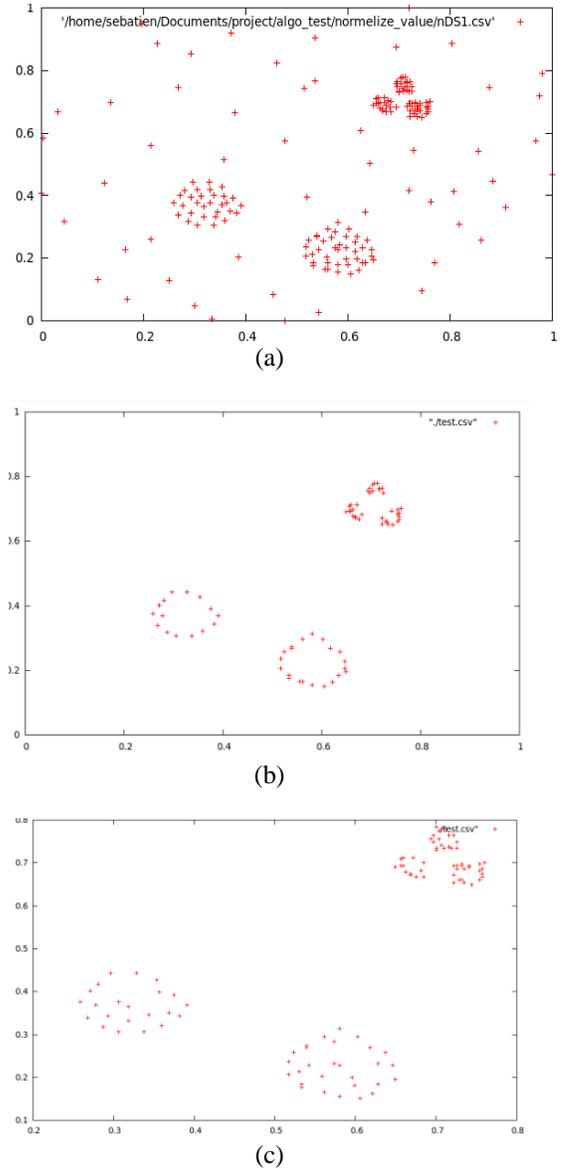

Figure 3. Boundary Detection with DS1.

Next, all local models will be sent to the server where we build the global model during phase 3. The construction of global model also includes two steps: merging local boundaries to create boundaries of global clusters and then regenerating data points to build global clusters. Fig.4 shows the whole dataset DS9 (Fig.4a) as well as its partitions on each node (Fig.4b, 4c, 4d). Fig.5 shows the boundaries detected of clusters found in each node. The global cluster is in Fig. 6 where we merge local boundaries and regenerate data points by random throw algorithm (cf. Table II). By observing the figures 4, 5, and 6, we note that:

- The regenerated clusters are similar to the original ones (Fig.4) in terms of overall shape.

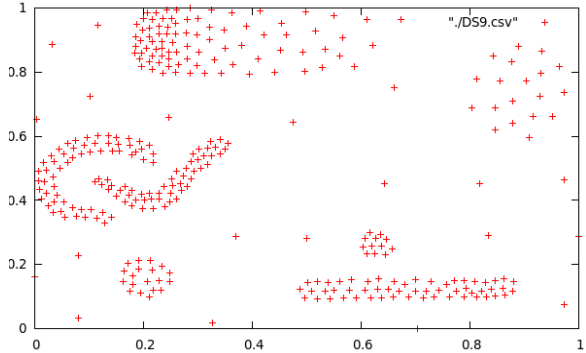

(a)

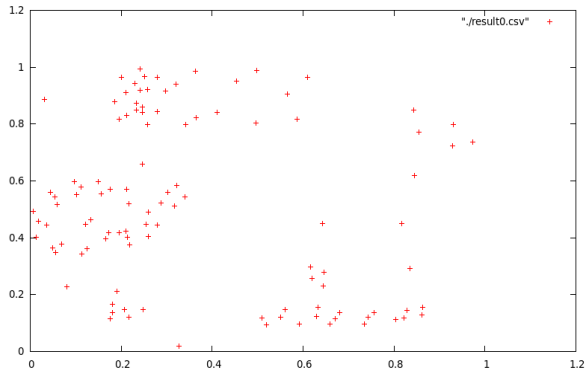

(b)

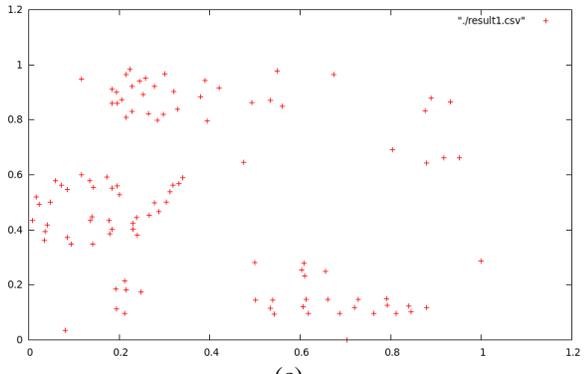

(c)

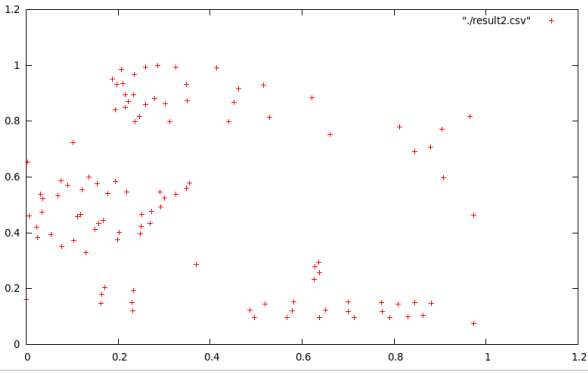

(d)

Figure 4. Dataset DS9 and its local partitions.

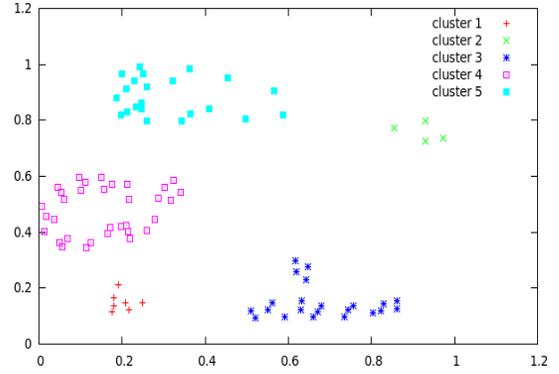

(a) Node1

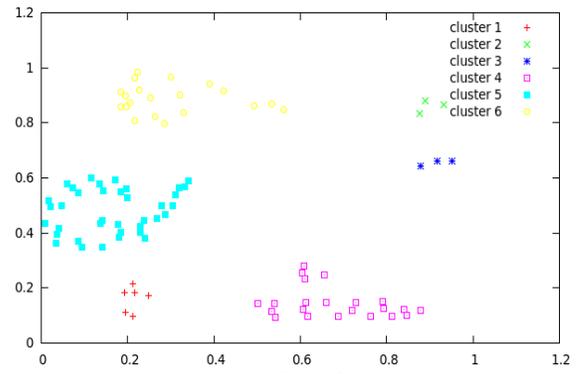

(b) Node 2

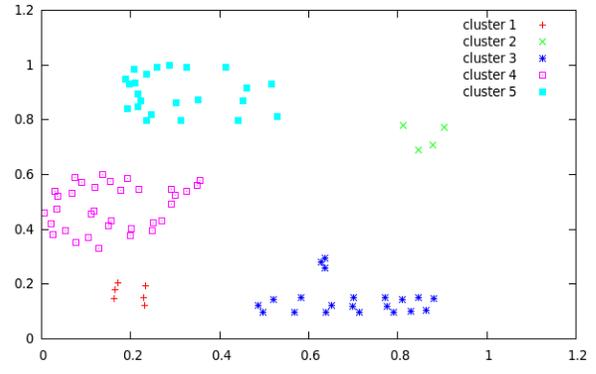

(c) Node 3

Figure 5. Clustering and Boundary Detection at local node.

- There is a slight difference in the density between the original datasets (Fig.4a). The reason is that in this approach we applied the cardinality of the cluster not its density to regenerate the clusters.

The idea behind this strategy is pretty simple but it has an issue. It often happens that some points that form small areas of relatively high density and small "bubbles" of empty space, may create a cluster C' of non-constant density around all of its points.

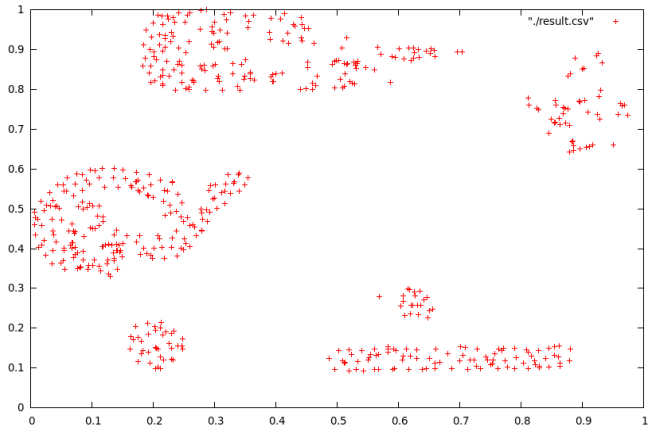

Figure 6. Merging.

## V. CONCLUSIONS AND FUTURE WORKS

In this paper, we present a new approach for distributed clustering techniques. In this approach, local models are not directly merged to build the global ones. The local models are extracted from the local datasets so that their sizes are small enough to send through the network. Besides, we regenerate local datasets from their local models and then merge them together to build the new global models that will be analysed by other mining techniques. Preliminary results of this algorithm are also presented and discussed. They also show the feasibility and usefulness of our approach. As we can see, this method is different from the existing distributed clustering models presented in the literature. Most of the current methods are based on aggregating local models to build the global ones.

A more extensive evaluation is on going. In the future we intend to analyse large real world datasets such as Hurricane data [16] in order to improve our algorithms as well as define efficient parameters to cope with complex shapes. Indeed, we are currently testing with different regeneration approaches such as grid strategy, perturbed grid strategy to cope with the difference in density of clusters.